\documentclass[conference]{IEEEtran}
\IEEEoverridecommandlockouts
\usepackage{cite}
\usepackage{amsmath,amssymb,amsfonts}
\usepackage{algorithm}
\usepackage{algorithmic}
\usepackage{graphicx}
\usepackage{textcomp}
\usepackage{xcolor}
\usepackage{balance}
\usepackage{multirow}
\usepackage{soul}

\usepackage{balance}
\def\BibTeX{{\rm B\kern-.05em{\sc i\kern-.025em b}\kern-.08em
    T\kern-.1667em\lower.7ex\hbox{E}\kern-.125emX}}
\begin{document}



\title{Black-Box Attacks against Signed Graph Analysis via Balance Poisoning}

\author{
    \IEEEauthorblockN{Jialong Zhou$^\dag$, Yuni Lai$^\dag$, Jian Ren$^*$, Kai Zhou$^\dag$}
    \IEEEauthorblockA{$^\dag$ Department of Computing, The Hong Kong Polytechnic University, HKSAR}
    \IEEEauthorblockA{$^*$ Dept of ECE, Michigan State University, Michigan, USA}
    \IEEEauthorblockA{\{jialong.zhou, yunie.lai\}@connect.polyu.hk,  renjian@msu.edu, kaizhou@polyu.edu.hk}
}

\maketitle

\begin{abstract}

Signed graphs are well-suited for modeling social networks as they capture both positive and negative relationships. Signed graph neural networks (SGNNs) are commonly employed to predict link signs (i.e., positive and negative) in such graphs due to their ability to handle the unique structure of signed graphs.
However, real-world signed graphs are vulnerable to malicious attacks by manipulating edge relationships, and existing adversarial graph attack methods do not consider the specific structure of signed graphs. SGNNs often incorporate balance theory to effectively model the positive and negative links. Surprisingly, we find that the balance theory that they rely on can ironically be exploited as a black-box attack. In this paper, we propose a novel black-box attack called \textit{balance}-attack that aims to decrease the balance degree of the signed graphs. We present an efficient heuristic algorithm to solve this NP-hard optimization problem. We conduct extensive experiments on five popular SGNN models and four real-world datasets to demonstrate the effectiveness and wide applicability of our proposed attack method. By addressing these challenges, our research contributes to a better understanding of the limitations and resilience of robust models when facing attacks on SGNNs. This work contributes to enhancing the security and reliability of signed graph analysis in social network modeling. Our PyTorch implementation of the attack is publicly available on GitHub: https://github.com/JialongZhou666/Balance-Attack.git.
\end{abstract}


\section{Introduction}

In social interactions, relationships can encompass both positive aspects, such as trust, and negative aspects, such as hate. Signed graphs provide a suitable network structure for capturing these diverse relationships. By incorporating positive and negative signs, signed graphs effectively represent both friendly and hostile connections, making them well-suited for modeling various social networks. An example of such a scenario is the Bitcoin-Alpha platform, where users can rate others using positive or negative scores. This natural setting can be effectively modeled using a signed graph. Machine learning techniques have played a significant role in analyzing signed graph data, addressing tasks such as link sign prediction \cite{leskovec2010predicting, song2015link} and node ranking \cite{tang2016survey, chen2018bridge}.


This paper primarily focuses on the task of link sign prediction, which involves inferring the signs of edges in the unexplored portion of a signed graph based on a known subgraph with its structure and edge signs. Numerous existing approaches for link sign prediction rely on signed graph neural networks (SGNNs) \cite{derr2018signed, huang2021signed, huang2021sdgnn, li2020learning, shu2021sgcl}. SGNNs are constructed based on graph neural networks (GNNs) \cite{velivckovic2017graph} and are specifically designed to accommodate the unique graph structure of signed graphs. Since the presence of negative edges invalidates the standard message-passing mechanism, the development of SGNN models becomes necessary to effectively handle both positive and negative edges.

To effectively address the challenges associated with negative edges in the design of SGNNs, a common approach is to incorporate balance theory from social psychology. This theory provides valuable insights into managing and integrating positive and negative links, thereby establishing a cohesive framework for learning node representations. Balance theory suggests the existence of an expected ``balanced structure" in which signed triangles, composed of three interconnected nodes, should have an even number of negative edges \cite{cartwright1956structural}. Empirical studies have confirmed that most triangles in real-world signed social graph datasets adhere to these conditions \cite{leskovec2010signed, leskovec2010predicting}. Existing SGNN models, such as Signed Graph Convolutional Networks (SGCN) \cite{derr2018signed} and Signed Network Embedding via graph Attention (SNEA) \cite{li2020learning}, have leveraged balance theory in the design of their aggregation strategies. 


However, real-world signed graphs are vulnerable to malicious attacks. For instance, in bitcoin trading platforms, users may engage in manipulative behavior by providing false ratings, while in e-commerce sites, attackers can disrupt the integrity of the award system by assigning low scores. These attacks typically involve altering a small portion of the edge relationships within the signed network. Such manipulations can have a significant impact on the results of link sign prediction using SGNNs, potentially leading to the deterioration of social relationships.

To understand the vulnerability of SGNNs, it is necessary to develop attack methods for signed graphs. Existing adversarial graph attack methods like Nettack \cite{bojchevski2019adversarial} and Metattack \cite{zugner_adversarial_2019} are not suitable, as they primarily require node labels and features for node classification tasks. Therefore, a new attack method tailored for signed graphs is required. Currently, there is a noticeable lack of black-box attacks for signed graphs. The only existing method that somewhat resembles a black-box attack is the random attack, where the signs of some edges are randomly altered. However, this method proves to be ineffective.

Given that most SGNN models rely on balance theory to aggregate information, either directly or implicitly, we propose a novel black-box attack for SGNNs by reducing the balance degree, which we termed as \textbf{\textit{balance}-attack}.
It has been proved that a signed graph neural network is incapable of learning accurate node representations from unbalanced triangles \cite{zhang2023rsgnn}.
By developing an algorithm to manipulate the degree of balance, our proposed \textbf{\textit{balance}-attack} shows to be effective. The major contributions of our research are as follows: 
\begin{itemize}
    \item We introduce a novel black-box attack for signed graph neural networks by corrupting the balance degree.
    \item We propose an effective and efficient algorithm to reduce the balance degree of signed graphs, a problem that has been proven to be NP-hard~\cite{diao2020approximation}.
    \item We conduct extensive experiments on four datasets using five popular SGNN models to demonstrate the effectiveness and generality of our proposed attack.
\end{itemize}
By addressing these issues, our aim is to advance the understanding of the limitations and resilience of robust models when faced with attacks on signed graph neural networks.

\section{Related Work}



Extensive research has been conducted in the machine learning and security communities to explore adversarial attacks across different types of models. While naturally occurring outliers in graphs present certain challenges, adversarial examples are intentionally crafted to deceive machine learning models with unnoticeable perturbations. GNNs are particularly susceptible to these small adversarial perturbations in the data. As a result, numerous studies have focused on investigating adversarial attacks specifically targeted at graph learning tasks. 
Bojchevski et al. \cite{bojchevski2019adversarial} propose poisoning attacks on unsupervised node representation learning or node embedding, leveraging perturbation theory to maximize the loss incurred after training DeepWalk. Zugner et al. \cite{zugner_adversarial_2019}, on the other hand, tackle the inherent bi-level problem in training-time attacks by employing meta-gradients, effectively treating the graph as a hyper-parameter to optimize.

However, it is important to highlight that aforementioned studies primarily focus on unsigned graphs. When it comes to signed graphs, there is limited research in the context of adversarial attacks. Godziszewski et al. \cite{godziszewski2021attacking} introduce the concept of attacking sign prediction, where an attacker aims to conceal the signs of a specific set of target links from a network analyst by eliminating the signs of non-target links. However, this method is not specifically designed for SGNN models, and it does not function as a black-box attack. To the best of our knowledge, there have been no reported instances of adversarial attacks specifically tailored to signed graphs in a black-box manner thus far.

\section{Preliminaries}
\subsection{Balance Theory}

\begin{figure}[]
\centerline{\includegraphics[width=0.42\textwidth]{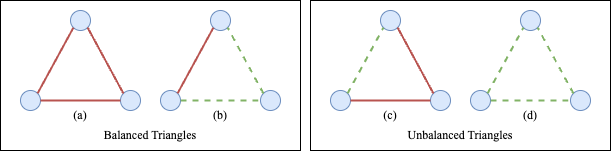}}
\caption{Four types of triangles.}
\label{triangle}
\end{figure}

In balance theory, balanced triads in a graph are defined as triads that contain an even number of negative edges. For instance, in Fig. \ref{triangle}, where positive and negative edges are represented by red solid and green dashed lines, respectively, the first two triads, where all three users are friends or only one pair of them are friends, are considered balanced. Based on previous works, this theory posits that individuals within a social network have a propensity to form structures that adhere to balance \cite{leskovec2010predicting, leskovec2010signed}. To quantify the degree of balance in a signed graph, a measurement called balance degree ($D_3(G)$) was introduced \cite{cao2015grarep}. It calculates the fraction of balanced triads in the graph using the following formula:
\begin{equation}
\label{eqn:BalanceDegree}
    D_3(G) = \frac{Tr(A^3)+Tr(|A|^3)}{2Tr(|A|^3)},
\end{equation}
where $Tr(\cdot)$ represents the trace of a matrix, $A$ is the signed adjacency matrix of the signed graph $G$. The elements in matrix $A$ can take values of $\{-1, 1, 0\}$ to represent negative edges, positive edges, or the absence of an edge in the signed graphs.


\subsection{Signed Graph Analysis}

Link sign prediction is a crucial task of analyzing signed graphs, as it entails deducing the signs of edges in the uncharted section of the graph. This prediction relies on a known subgraph, encompassing both its structure and edge signs. In the realm of signed graph analysis, link sign prediction takes precedence over other tasks such as node ranking.

SGCN \cite{derr2018signed}, the pioneering SGNN model, extends GCN to handle signed graphs by incorporating balance theory to determine positive and negative relationships between nodes. To provide further clarity, the representation of a node $v_i$ at a given layer $l$ is defined as:
\begin{equation}
    h_i^{(l)} = [h_i^{pos(l)}, h_i^{neg(l)}],
\end{equation}
where $h_i^{pos(l)}$ and $h_i^{neg(l)}$ respectively denote the positive and negative representation vectors of node $v_i \in \mathcal{V}$ at the $l$th layer, and $[\cdot,\cdot]$ denotes the concatenation operation. The updating process for $l>1$ layer could be written as:
\begin{equation}
\begin{split}
    & t_i^{pos(l)}=AGG^{(l)}({h_j^{pos(l-1)}:v_j \in \mathcal{N}_i^+},{h_j^{neg(l-1)}:v_j \in \mathcal{N}_i^-}) \\
    & h_i^{pos(l)}=COM^{(l)}(h_i^{pos(l-1)},t_i^{pos(l)}) \\
    & t_i^{neg(l)}=AGG^{(l)}({h_j^{neg(l-1)}:v_j \in \mathcal{N}_i^+},{h_j^{pos(l-1)}:v_j \in \mathcal{N}_i^-}) \\
    & h_i^{neg(l)}=COM^{(l)}(h_i^{neg(l-1)},t_i^{neg(l)}),
\end{split}
\end{equation}
where $AGG$ and $COM$ refers to the aggregation and combination processes, respectively. $t_i$ represents the temporary node representation vectors after the aggregation step. The set $\mathcal{N}$ corresponds to the neighbors of node $v_i$. SGNNs handle positive and negative edges by employing a two-part representation and a unique aggregation scheme. For instance, when $l>1$, the positive part of the representation for node $v_i$ aggregates information from the positive representations of its positive neighbors and the negative representations of its negative neighbors. In other SGNN models, they hold similar mechanism. SGCL \cite{shu2021sgcl} and UGCL \cite{ko2023universal} utilize graph contrastive learning for signed graphs.. SGDNN \cite{huang2021sdgnn} combines balance theory and status theory along with the introduction of four weight matrices. RSGNN \cite{zhang2023rsgnn} incorporates structure-based regularizers to enhance performance.

\section{Problem Statements}
We focus on the task of link sign prediction, which involves predicting the signs of edges in the complementary part of a given subgraph of a signed graph with known structure and edge signs. To begin, we introduce the necessary notations and formulate the attack model accordingly.

\subsection{Notations}
Formally, let $G=(V,E^+,E^-)$ be a signed-directed graph where $V=\{v_1,v_2, \cdots,v_n\}$ represents the set of $n$ nodes. The positive edges are denoted by $E^+ \subseteq V \times V$, while the negative edges are $E^- \subseteq V \times V$, and $E^+ \cap E^- = \emptyset$. We denote the sign of edge $e_{ij}$ as $\sigma (e_{ij}) \in \{ +, -\}$. The structure of $G$ is captured by the adjacency matrix $A \in \mathbb{R}^{|V| \times |V|}$, where each entry $A_{ij} \in \{1, -1, 0\}$  represent negative edges, positive edges, or the absence of an edge in the signed graphs. We denote the training edges and testing edges by $\mathcal{D}_{train}$ and $\mathcal{D}_{test}$, respectively, and each edge $e\in \mathcal{D}_{train} \cup \mathcal{D}_{test}$ has its sign label $\sigma(e)$. Let $\mathcal{L}_{train}$ be the training loss of the target model based on $\mathcal{D}_{train}$, and $\theta$ denote the model parameter. The model predictions for the sign of edges are denoted as $f_{\theta ^ *}(G)$, and $f_{\theta ^ *}(G)_e \in \{+,-\}$ is the prediction for the given edge $e \in E^+ \cap E^-$. $\mathcal{L}_{train}$ is the training loss of the target model and $\mathcal{L}_{atk}$ represents the objective that the attacker seeks to optimize. Let $\mathbb{I}\{\cdot\}$ be the indicator function, and $sign(\cdot)$ be the sign function.

\subsection{Threat Model}



\subsubsection{Attacker's goal}
Our study aims to investigate the vulnerability of link sign prediction models by developing a black-box attack that aims to assess the extent to which the predictions of the algorithm can be disturbed. Following \cite{zugner_adversarial_2019}, we focus on global attacks, aiming to decrease the overall prediction performance of the model. We leverage an attack method to manipulate the graph effectively. The modified graph is then utilized to train SGNNs, intentionally aiming to degrade their performance.


\subsubsection{Attacker's knowledge}
We assume that the attackers have access to the training data, enabling them to observe both the graph structure and edge signs, but they do not know the model structure and parameters.

\subsubsection{Attacker's capability}
To ensure effective and inconspicuous adversarial attacks, we impose a budget constraint denoted as $\Delta$, limiting the number of changes made to the graph. Specifically, the constraint restricts the number of altered edges $\lVert A - \hat{A} \lVert_{0}$ to stay within $\Delta$. In our case, we disregard changes in edge signs and assume graph symmetry, resulting in a budget constraint of $2\Delta$. We also take precautions to prevent node disconnection during the attack process. Unnoticeability of changes is maintained by imposing a constraint on the degree distribution. Although our current focus is altering edge signs, our algorithm can be easily adapted to modify the overall graph structure. These constraints are consolidated as the set of permissible perturbations on the given graph $G$, denoted as $\Phi(G;\Delta)$.


\subsection{Problem of Attack}
In the case of global and unspecific attacks, the primary aim of the attacker is to reduce the model's generalization performance on the testing nodes. Poisoning attacks can be mathematically formulated as a bi-level optimization problem:
\begin{align}
\label{eqn:BilevelOptimization}
    & \mathop{\min}_{\hat{G} \in \Phi(G;\Delta)} \mathcal{L}_{atk} = \sum_{e \in \mathcal{D}_{test}}\mathbb{I}\{f_{\theta ^ *}(\hat{G})_e=\sigma(e)\}, \\ 
    &s.t. \  \theta^* = \mathop{arg \min}_{\theta} \mathcal{L}_{train}(f_\theta (\hat{G})),\nonumber
\end{align}
where the attacker aims to reduce the number of testing edges to be correctly classified by manipulating the graph, and the model itself is trained on the manipulated graph.


\section{Proposed Black-Box Attack}

\subsection{Formulation of black-box attack}
Since the model structure and labels of the testing data are always unavailable, directly optimizing \eqref{eqn:BilevelOptimization} becomes infeasible. To address this challenge, we adopt an alternative approach by minimizing the balance degree of the graph.
According to the analysis conducted in a previous study \cite{zhang2023rsgnn}, it has been determined that SGNNs lack the ability to effectively learn precise node representations from unbalanced triangles. From this finding, we can infer that targeting the balance attribute of graphs has the potential to degrade the performance of SGNNs. Consequently, if the target model $\theta$ is trained on a poisoned graph that has a low balance degree, it is expected to exhibit an also low $\mathcal{L}_{atk}$ value.
Therefore, we replace the optimization problem \eqref{eqn:BilevelOptimization} with the optimization problem as follows:
\begin{equation}
\label{eqn:BalanceOptimization}
    \mathop{\min}_{\hat{G} \in \Phi(G;\Delta)} D_3(\hat{G}).
\end{equation}






\subsection{Attack Method}


In the training phase, our objective is to minimize the balance degree of the subgraph $\hat{G}$ within a specified budget $\Delta$. This problem, however, is challenging due to the discrete nature of the signs. As mentioned in \cite{diao2020approximation}, optimizing this problem is known to be NP-hard. To approximate the optimization problem, we propose an algorithm based on gradient descent and greedy edge selection.




Our solution revolves around the core concept of computing the gradient of the objective function $D_3(\hat{G})$ with respect to the adjacency matrix $A$. The main strategy is to iteratively and greedily flip the sign of an existing edge that has the largest absolute gradient value and the correct sign, while ensuring adherence to the budget constraint. This process is repeated until the budget is fully consumed. During each iteration, we update an element in the adjacency matrix using the following procedure:
\begin{equation}
\label{eqn:GreedyEdgeSignUpdate}
\begin{aligned}
    & i^*,j^* = \mathop{arg \max}_{\{i,j|a_{ij}\neq 0 \wedge sign(a_{i,j}) = \atop sign(\nabla _{ij}D_3(\hat{G}))\}} |\nabla _{ij} D_3(\hat{G})|, \\
    & a_{i,j} = -a_{i^*,j^*},
\end{aligned}
\end{equation}
where $a_{i,j}$ represents an element located at row $i$ and column $j$ of the adjacency matrix. The variable $\nabla _{ij} D_3(\hat{G})$ denotes the gradient of each edge computed through back-propagation. When selecting $i^*$ and $j^*$, it is crucial to ensure that the signs of the gradient and the edge are the same. To provide a clearer understanding of our approach, we outline the steps of our greedy flips method in Alg.~\ref{alg:1}.
\begin{algorithm}
\label{alg:1}
    \caption{Algorithm of \textbf{\textit{balance}-attack} via Greedy Flips}
    \renewcommand{\algorithmicrequire}{\textbf{Input:}}
    \renewcommand{\algorithmicensure}{\textbf{Output:}}
    \begin{algorithmic}[1]
        \REQUIRE Adjacency matrix $A$ of $G$, perturbation budget $\Delta$.
        \ENSURE Attacked adjacency matrix $S$ ($s$ is the element in $S$).
        \STATE Initialize $S\gets A$.
        \WHILE {Number of changed edges $\le$ $\Delta$}
        \STATE Calculate $D_3(S)$.
        \STATE Calculate gradient matrix $\nabla(D_3(S))$.
        \STATE Filter candidate edges $C_e=\{i,j|s_{ij}\neq 0 \wedge  sign(s_{i,j}) = sign(\nabla _{ij}D_3(S))\}$.
        \IF {$i^*,j^* = \mathop{arg \max}_{\{i,j\in C_e\}} |\nabla _{ij}(D_3(S))|$}
        \STATE Update $s_{i^*,j^*}=-s_{i^*,j^*}$.
        \STATE Number of changed edges $+=1$.
        \ENDIF
        \ENDWHILE
        \STATE Return $S$.
    \end{algorithmic}
\end{algorithm}

\section{Experiments}

In this section, we perform experiments on \textit{4} real-world datasets to showcase the efficacy of the proposed \textbf{\textit{balance}-attack} in diminishing the performance of SGNNs compared to random attacks in link sign prediction. Additionally, we apply \textbf{\textit{balance}-attack} to \textit{5} state-of-the-art methods in signed graph representation. We will answer the following questions:

\begin{itemize}
\item \textbf{Q1}: Can \textbf{\textit{balance}-attack} decrease the balance degree of signed graphs significantly?
\item \textbf{Q2}: How does \textbf{\textit{balance}-attack} perform on existing SGNN models compared with random attack?
\item \textbf{Q3}: How applicable is \textbf{\textit{balance}-attack} on various SGNN models?
\end{itemize}

\begin{table}[]
\caption{Dataset Statistics}
\setlength{\tabcolsep}{2.2pt}
\begin{tabular}{cccccc}
\hline
\textbf{Dataset}       & \textbf{\#Nodes} & \textbf{\#Pos-Edges} & \textbf{\#Neg-Edges} & \textbf{\%Pos-Ratio} & \textbf{\%Density} \\ \hline
Bitcoin-Alpha & 3,784   & 22,650       & 1,536        & 93.65        & 0.3379\%   \\
Bitcoin-OTC   & 5,901   & 32,029       & 3,563        & 89.99        & 0.2045\%   \\
Slashdot      & 33,586  & 295,201      & 100,802      & 74.55        & 0.0702\%   \\
Epinions      & 16,992  & 276,309      & 50,918       & 84.43        & 0.2266\%   \\ \hline
\end{tabular}
\label{dataset}
\end{table}

\subsection{Baseline}

To establish a baseline for comparison, we employ a random attack strategy since there is currently no established black-box attack model specifically designed for signed graphs. Existing adversarial graph attack methods such as Metattack \cite{zugner_adversarial_2019} and CLGA \cite{zhang2022unsupervised} are primarily developed for unsigned graphs. While these methods can be adapted for signed graphs, they heavily rely on node labels and node features, making them unsuitable for this particular scenario. In the case of the random attack, we randomly select a set of edges from the input signed graph and flip their signs. 

\begin{figure}[]
\centerline{\includegraphics[width=0.35\textwidth]{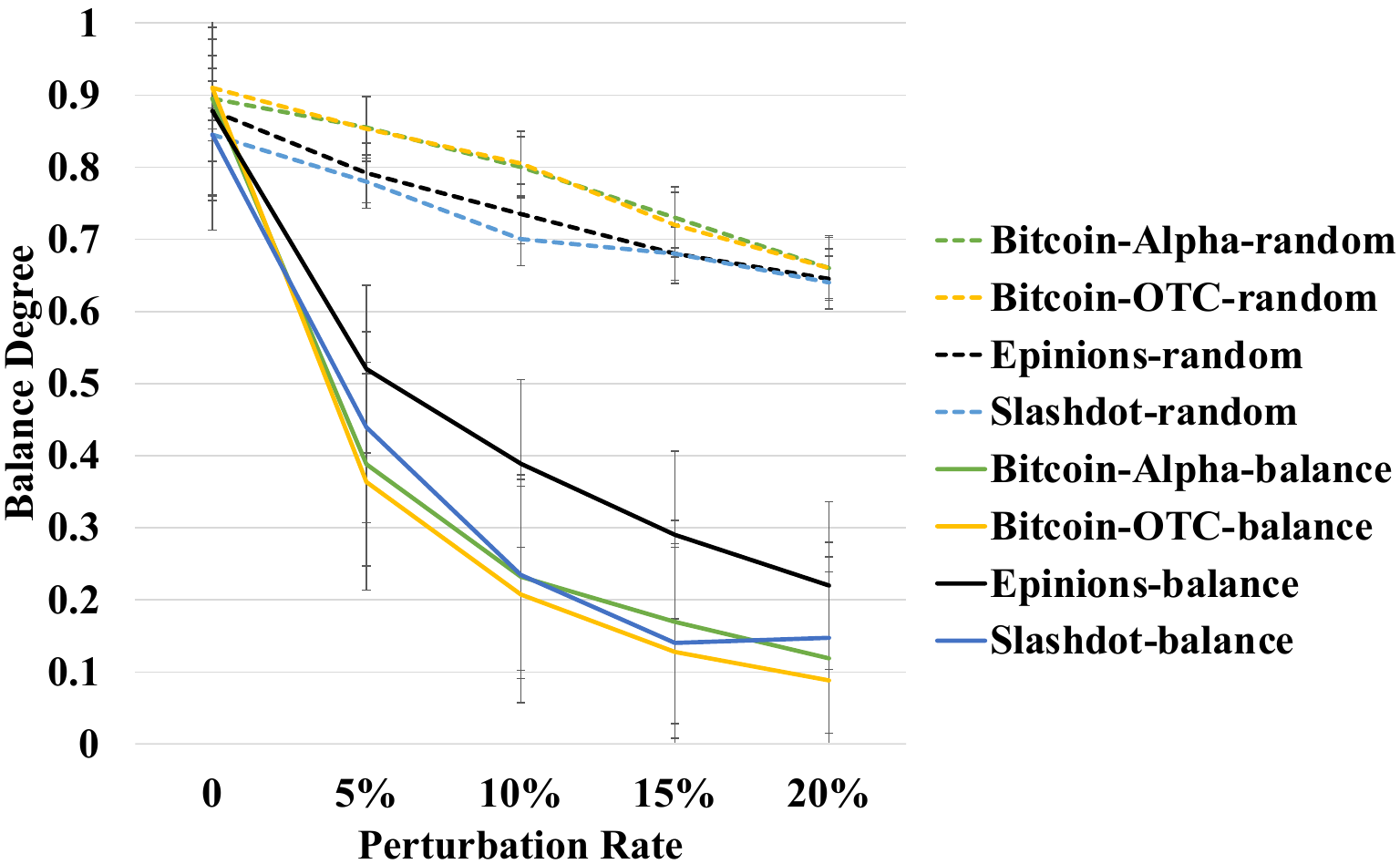}}
\caption{Balance degree of $4$ datasets under balance-attack and random attack.}
\label{balance_degree}
\end{figure}

\subsection{Setup}

We conduct experiments on four public real-world datasets: Bitcoin-Alpha, Bitcoin-OTC \cite{kumar2016edge}, Epinions \cite{guha2004propagation}, and Slashdot \cite{guo2017hermitian}. The Bitcoin-Alpha and Bitcoin-OTC datasets are publicly available and collected from Bitcoin trading platforms. These datasets are obtained from platforms where users have the ability to label other users as either trust (positive) or distrust (negative) users. This labeling system serves as a means to prevent transactions with fraudulent and risky users from trading or perform transactions, given the anonymity of these trading platforms. Slashdot is a renowned technology-related news website that boasts a distinctive user community. Within this community, users have the option to tag each other as friends or foes based on their interactions and relationships. Similarly, Epinions represents an online social network centered around a general consumer review site called Epinions.com. The users of this site have the autonomy to decide whether they trust other members or not, forming a network based on mutual trust relationships. 
In the experiments, we randomly select $80\%$ links as training set and the remaining $20\%$ as testing set. Since these datasets have no attributes, we randomly generate a $64$-dimensional vector for each node as the initial node attribute. More detailed dataset statistics are shown in Table \ref{dataset}.

With the above benchmark datasets, we evaluate \textbf{\textit{balance}-attack} on five popular SGNN models, as follows:
\begin{itemize}
\item \textbf{SGCN} \cite{derr2018signed} aims to bridge the gap between unsigned GCN and the analysis of signed graphs. It strives to develop a novel information aggregator by leveraging balance theory, thereby extending the applicability of GCN to signed graphs.
\item \textbf{SGCL} \cite{shu2021sgcl} is the first work to generalize graph contrastive learning to signed graphs, which employs graph augmentations to reduce the harm of noisy interactions and enhances the model robustness.
\item \textbf{SGDNN} \cite{huang2021sdgnn} combines both balance theory and status theory, and introduces four weight matrices to aggregate neighbor features based on edge types.
\item \textbf{RSGNN} \cite{zhang2023rsgnn} incorporates structure-based regularizers to enhance the performance of SGNNs by emphasizing the intrinsic properties of a signed graph and mitigating their vulnerability to potential edge noise in the input graph. This noise-tolerant SGNN model leverages the regularization techniques to improve the robustness and reliability of the model against noisy and unreliable edges in the graph.
\item \textbf{UGCL} \cite{ko2023universal} introduces a novel contrastive learning framework that incorporates Laplacian perturbation, offering a unique advantage through the utilization of an indirect perturbation method that ensures stability and maintains effective perturbation effects.
\end{itemize}

\begin{table}[t] \centering
\caption{Link sign prediction performance of RSGNN under random attack and \textbf{\textit{Balance}-attack}}
\setlength{\tabcolsep}{3pt}
\begin{tabular}{|c|c|c|ccc|}
\hline
Dataset                        & Ptb                   & Attack                    & Micro\_f1       & Binary\_f1      & Macro\_f1       \\ \hline
\multirow{9}{*}{Bitcoin-Alpha} & 0                     & -                         & 0.8820           & 0.9341          & 0.6841          \\ \cline{2-6} 
                               & \multirow{2}{*}{5\%}  & random                    & 0.8299          & 0.9018          & 0.6317          \\ \cline{3-3}
                               &                       & \textit{\textbf{balance}} & \textbf{0.7563} & \textbf{0.8534} & \textbf{0.5648} \\ \cline{2-6} 
                               & \multirow{2}{*}{10\%} & ranodm                    & 0.7726          & 0.8642          & 0.5831          \\ \cline{3-3}
                               &                       & \textit{\textbf{balance}} & \textbf{0.6802} & \textbf{0.7984} & \textbf{0.5123} \\ \cline{2-6} 
                               & \multirow{2}{*}{15\%} & random                    & 0.7360           & 0.8393          & 0.5499          \\ \cline{3-3}
                               &                       & \textit{\textbf{balance}} & \textbf{0.6605} & \textbf{0.7862} & \textbf{0.4814} \\ \cline{2-6} 
                               & \multirow{2}{*}{20\%} & random                    & 0.6839          & 0.8010           & 0.5165          \\ \cline{3-3}
                               &                       & \textit{\textbf{balance}} & \textbf{0.6308} & \textbf{0.7631} & \textbf{0.4635} \\ \hline
\multirow{9}{*}{Bitcoin-OTC}   & 0                     & -                         & 0.8919          & 0.9382          & 0.7553          \\ \cline{2-6} 
                               & \multirow{2}{*}{5\%}  & random                    & 0.8654          & 0.9216          & 0.7227          \\ \cline{3-3}
                               &                       & \textit{\textbf{balance}} & \textbf{0.7970}  & \textbf{0.8761} & \textbf{0.6574} \\ \cline{2-6} 
                               & \multirow{2}{*}{10\%} & random                    & 0.8242          & 0.8950           & 0.6782          \\ \cline{3-3}
                               &                       & \textit{\textbf{balance}} & \textbf{0.7134} & \textbf{0.8158} & \textbf{0.5849} \\ \cline{2-6} 
                               & \multirow{2}{*}{15\%} & random                    & 0.8180           & 0.8911          & 0.6694          \\ \cline{3-3}
                               &                       & \textit{\textbf{balance}} & \textbf{0.6787} & \textbf{0.7909} & \textbf{0.5486} \\ \cline{2-6} 
                               & \multirow{2}{*}{20\%} & random                    & 0.7828          & 0.8673          & 0.6341          \\ \cline{3-3}
                               &                       & \textit{\textbf{balance}} & \textbf{0.6424} & \textbf{0.7625} & \textbf{0.5194} \\ \hline
\multirow{9}{*}{Slashdot}      & 0                     & -                         & 0.7823          & 0.8574          & 0.6988          \\ \cline{2-6} 
                               & \multirow{2}{*}{5\%}  & random                    & 0.7384          & 0.8321          & 0.6629          \\ \cline{3-3}
                               &                       & \textit{\textbf{balance}} & \textbf{0.7344} & \textbf{0.8225} & \textbf{0.6484} \\ \cline{2-6} 
                               & \multirow{2}{*}{10\%} & random                    & 0.7092          & 0.7982          & 0.639           \\ \cline{3-3}
                               &                       & \textit{\textbf{balance}} & \textbf{0.6719} & \textbf{0.7761} & \textbf{0.5813} \\ \cline{2-6} 
                               & \multirow{2}{*}{15\%} & random                    & 0.6707          & 0.7637          & 0.6104          \\ \cline{3-3}
                               &                       & \textit{\textbf{balance}} & \textbf{0.6378} & \textbf{0.7466} & \textbf{0.5557} \\ \cline{2-6} 
                               & \multirow{2}{*}{20\%} & random                    & 0.6637          & 0.7576          & 0.6044          \\ \cline{3-3}
                               &                       & \textit{\textbf{balance}} & \textbf{0.6009} & \textbf{0.7165} & \textbf{0.5215} \\ \hline
\multirow{9}{*}{Epinions}      & 0                     & -                         & 0.8280           & 0.8932          & 0.7261          \\ \cline{2-6} 
                               & \multirow{2}{*}{5\%}  & random                    & 0.8155          & 0.8841          & 0.7160           \\ \cline{3-3}
                               &                       & \textit{\textbf{balance}} & \textbf{0.7736} & \textbf{0.8542} & \textbf{0.6739} \\ \cline{2-6} 
                               & \multirow{2}{*}{10\%} & random                    & 0.7711          & 0.8516          & 0.6754          \\ \cline{3-3}
                               &                       & \textit{\textbf{balance}} & \textbf{0.7342} & \textbf{0.8234} & \textbf{0.6432} \\ \cline{2-6} 
                               & \multirow{2}{*}{15\%} & random                    & 0.7376          & 0.8257          & 0.6475          \\ \cline{3-3}
                               &                       & \textit{\textbf{balance}} & \textbf{0.7068} & \textbf{0.8016} & \textbf{0.6201} \\ \cline{2-6} 
                               & \multirow{2}{*}{20\%} & random                    & 0.7409          & 0.8285          & 0.6492          \\ \cline{3-3}
                               &                       & \textit{\textbf{balance}} & \textbf{0.6832} & \textbf{0.7836} & \textbf{0.5962} \\ \hline
\end{tabular}
\label{RSGNN}
\end{table}

We follow the hyper-parameter setting suggestions by those papers and set the embedding dimension to $64$ for all SGNN models to achieve a fair comparison. To speed up the attack process, we opt to modify 10 edges per epoch. Specifically, we target the 10 elements in the adjacency matrix that possess the highest absolute gradient values and the correct signs, when doing back-propagation. In the experiment, the perturbation rate varies from $5\%$ to $20\%$ of total edges. To evaluate our method, we employ three metrics: micro-average F1 score (Micro-F1), binary-average F1 score (Binary-F1), and macro-average F1 score (Macro-F1). These metrics have been widely used in previous studies and provide valuable insights into the performance of SGNN models. Lower values of these metrics indicate poorer model performance and greater effectiveness of attack methods. However, we find that the area under the curve (AUC) metric may not be suitable for assessing the performance of models on signed graph datasets. AUC tends to yield misleading results on imbalanced datasets, which is the case for signed graph datasets that predominantly contain positive edges. Therefore, we exclude the AUC metric from our evaluation.

\begin{table}[t] \centering
\caption{Link sign prediction performance of SGNNs under random attack and \textbf{\textit{Balance}-attack} with perturbation rate = $20\%$}
\setlength{\tabcolsep}{3pt}
\begin{tabular}{|c|c|c|ccc|}
\hline
Model                  & Dataset                        & Attack                    & Micro\_F1       & Binary\_F1      & Macro\_F1       \\ \hline
\multirow{8}{*}{UGCL}  & \multirow{2}{*}{Bitcoin-Alpha} & random                    & 0.9199          & 0.9576          & 0.6192          \\ \cline{3-3}
                       &                                & \textit{\textbf{balance}} & \textbf{0.8044} & \textbf{0.8883} & \textbf{0.5526} \\ \cline{2-6} 
                       & \multirow{2}{*}{Bitcoin-OTC}   & random                    & 0.8988          & 0.9442          & 0.6983          \\ \cline{3-3}
                       &                                & \textit{\textbf{balance}} & \textbf{0.7752} & \textbf{0.8643} & \textbf{0.6044} \\ \cline{2-6} 
                       & \multirow{2}{*}{Slashdot}      & random                    & 0.8538          & 0.9173          & 0.6318          \\ \cline{3-3}
                       &                                & \textit{\textbf{balance}} & \textbf{0.7826} & \textbf{0.8704} & \textbf{0.5971} \\ \cline{2-6} 
                       & \multirow{2}{*}{Epinions}      & random                    & 0.8635          & 0.9237          & \textbf{0.6390}  \\ \cline{3-3}
                       &                                & \textit{\textbf{balance}} & \textbf{0.8328} & \textbf{0.9018} & 0.6665          \\ \hline
\multirow{8}{*}{SGCL}  & \multirow{2}{*}{Bitcoin-Alpha} & random                    & 0.9305          & 0.9636          & 0.6007          \\ \cline{3-3}
                       &                                & \textit{\textbf{balance}} & \textbf{0.8108} & \textbf{0.8931} & \textbf{0.5312} \\ \cline{2-6} 
                       & \multirow{2}{*}{Bitcoin-OTC}   & random                    & 0.9026          & 0.9480           & 0.6131          \\ \cline{3-3}
                       &                                & \textit{\textbf{balance}} & \textbf{0.7931} & \textbf{0.8785} & \textbf{0.5919} \\ \cline{2-6} 
                       & \multirow{2}{*}{Slashdot}      & random                    & 0.8338          & 0.9072          & 0.5578          \\ \cline{3-3}
                       &                                & \textit{\textbf{balance}} & \textbf{0.7002} & \textbf{0.8163} & \textbf{0.5001} \\ \cline{2-6} 
                       & \multirow{2}{*}{Epinions}      & random                    & 0.8482          & 0.9160           & \textbf{0.5673} \\ \cline{3-3}
                       &                                & \textit{\textbf{balance}} & \textbf{0.7385} & \textbf{0.8371} & 0.5872          \\ \hline
\multirow{8}{*}{SDGNN} & \multirow{2}{*}{Bitcoin-Alpha} & random                    & 0.8616          & 0.9234          & 0.6062          \\ \cline{3-3}
                       &                                & \textit{\textbf{balance}} & \textbf{0.7775} & \textbf{0.8698} & \textbf{0.5528} \\ \cline{2-6} 
                       & \multirow{2}{*}{Bitcoin-OTC}   & random                    & 0.8333          & 0.9028          & 0.6593          \\ \cline{3-3}
                       &                                & \textit{\textbf{balance}} & \textbf{0.7388} & \textbf{0.8371} & \textbf{0.5893} \\ \cline{2-6} 
                       & \multirow{2}{*}{Slashdot}      & random                    & 0.8405          & 0.8981          & 0.6966          \\ \cline{3-3}
                       &                                & \textit{\textbf{balance}} & \textbf{0.7326} & \textbf{0.8286} & \textbf{0.6106} \\ \cline{2-6} 
                       & \multirow{2}{*}{Epinions}      & random                    & 0.8336          & 0.9023          & 0.6714          \\ \cline{3-3}
                       &                                & \textit{\textbf{balance}} & \textbf{0.7696} & \textbf{0.8550}  & \textbf{0.6467} \\ \hline
\multirow{8}{*}{SGCN}  & \multirow{2}{*}{Bitcoin-Alpha} & random                    & 0.6614          & 0.7842          & 0.4991          \\ \cline{3-3}
                       &                                & \textit{\textbf{balance}} & \textbf{0.6022} & \textbf{0.7346} & \textbf{0.4704} \\ \cline{2-6} 
                       & \multirow{2}{*}{Bitcoin-OTC}   & random                    & 0.6833          & 0.7925          & 0.5620           \\ \cline{3-3}
                       &                                & \textit{\textbf{balance}} & \textbf{0.6265} & \textbf{0.7434} & \textbf{0.5288} \\ \cline{2-6} 
                       & \multirow{2}{*}{Slashdot}      & random                    & 0.6835          & 0.7752          & 0.6204          \\ \cline{3-3}
                       &                                & \textit{\textbf{balance}} & \textbf{0.5939} & \textbf{0.7029} & \textbf{0.5307} \\ \cline{2-6} 
                       & \multirow{2}{*}{Epinions}      & random                    & 0.6725          & 0.7712          & 0.5977          \\ \cline{3-3}
                       &                                & \textit{\textbf{balance}} & \textbf{0.6453} & \textbf{0.7497} & \textbf{0.5706} \\ \hline
\end{tabular}
\label{other_models}
\end{table}

\subsection{Balance Degree of Signed Graphs after Attack (Q1)}

To validate the effectiveness of our method, we first apply our approach and obtain conclusive results: the balance degree of signed graphs is significantly reduced compared to the balance degree under the random attack. We present the comparison results of the \textbf{\textit{balance}-attack} and random attack in Fig. \ref{balance_degree}. Initially, in each dataset, the balance degree ranges from $0.85$ to $0.9$. When subjected to random attacks with a perturbation rate of $20\%$, the minimum balance degree drops to approximately $0.65$. However, by utilizing our designed balance-attack method with a perturbation rate of $5\%$, the balance degree becomes more lower, ranging between $0.35$ and $0.55$. Furthermore, at a perturbation rate of $20\%$, the balance degree can be further reduced to about $0.1$, which is significantly lower than what is achieved through random attacks. These results unequivocally demonstrate the effectiveness of our proposed method in significantly reducing the balance degree of the graph.

\subsection{Attack Performance of Balance-Attack (Q2)}
We conduct a comparative analysis between random attack and \textbf{\textit{balance}-attack} on five existing SGNN models. To evaluate their performance, we tested the models at perturbation rates from $0\%$ to $20\%$ based on the three metrics mentioned before to evaluate the attack performance. RSGNN is a model known for its resilience against random attacks. While its original design may not have explicitly focused on adversarial attacks, we can infer that it possesses greater robustness against various attack scenarios compared to other SGNN models. Based on the results in Table \ref{RSGNN}, it is evident that RSGNN can maintain satisfactory performance even when subjected to random attacks. However, when exposed to our \textbf{\textit{balance}-attack}, the performance of RSGNN experiences a significant decline.
Similar results are observed in the other four SGNN models as presented in Table~\ref{other_models}.


\subsection{Applicability of Balance-Attack on various SGNNs (Q3)}
In addition to assessing balance-theory-based models (i.e., SGCN, SGDNN, and RSGNN), we also evaluate the performance of our proposed attack on non-balance-based models (i.e., SGCL and UGCL). It is important to note that, even though our attack method is designed based on the intuition that many SGNNs rely on balance theory, we surprisingly find that it also proves to be effective against non-balance-based SGNNs (Table \ref{RSGNN} and Table~\ref{other_models}). This showcases the versatility and efficacy of \textbf{\textit{balance}-attack} across different SGNNs.


\section{Conclusion}

In this paper, we introduce \textbf{\textit{balance}-attack}, a novel black-box attack for signed graphs, which reduces the balance degree. We propose an efficient heuristic algorithm to solve this NP-hard problem. Extensive experiments are conducted using popular SGNN models to validate the attack's effectiveness and generality. Our research aims to enhance the understanding of the limitations and resilience of robust models when faced with attacks on SGNNs.



\bibliographystyle{ieeetr}	
\bibliography{references}
\end{document}